\documentclass[a4paper,10pt,twoside]{cpc-hepnp}

\usepackage{multicol}
\usepackage{booktabs}
\usepackage{amssymb,mathrsfs,amscd}
\usepackage[tbtags]{amsmath}
\usepackage{lastpage}
\usepackage{braket}
\usepackage{slashed}
\usepackage{mathrsfs}
\usepackage{graphicx}
\usepackage{subfigure}
\usepackage{latexsym,bm}
\usepackage{multirow}
\usepackage{dcolumn}
\usepackage{CJK}
\newcommand{\MeV}{\text{MeV}}

\begin{document}
\newcolumntype{d}[1]{D{.}{.}{#1}}
%\begin{CJK*}{GBK}{song}
%\CJKtilde

%\date{\today}
\providecommand{\url}[1]{}
\bibliographystyle{hepnp}
\fancyhead[co]{\footnotesize Study the Heavy Molecular States in Quark Model with Meson
  Exchange Interaction}
\footnotetext[0]{Received \today}

\title{Study of the heavy molecular states in quark model with meson
  exchange Interaction}

\author{
   YU Si-Hai$^1$%
  \quad WANG Bao-Kai$^1$%
  \quad CHEN Xiao-Lin$^2$%
  \quad DENG Wei-Zhen$^{1;1)}$\email{dwz@pku.edu.cn}}
\maketitle
\address{%
  1~(Department of Physics and State Key
  Laboratory of Nuclear Physics and Technology, \\
  Peking University, Beijing 100871, China)\\
  2~(Department of Physics, Peking University, Beijing 100871, China)\\
}

\begin{abstract}
  Some charmonium-like resonances such as $X(3872)$ can be interpreted
  as possible $D^{(*)}\bar{D}^{(*)}$ molecular states.  Within the quark
  model, we study the structure of such molecular states and the
  similar $B^{(*)}\bar{B}^{(*)}$ molecular states by taking into account
  the light meson exchange ($\pi$, $\eta$, $\rho$, $\omega$ and
  $\sigma$) between two light quarks from different mesons.
\end{abstract}


\begin{multicols}{2}

\section{Introduction}

Some recently discovered narrow charmonium-like resonances have
aroused great theoretical interest.  The typical $X(3872)$ state,
which was discovered by the Belle Collaboration \cite{Choi:2003ue} and
subsequently confirmed by the CDF Collaboration \cite{Acosta:2003zx}
and BABAR Collaboration \cite{Aubert:2004ns}, is summarized
$M_X=3872.2\pm0.8\MeV$, $\Gamma_X=3.0^{+1.9}_{-1.4}\pm0.9\MeV$
\cite{Amsler:2008zzb}. Previously its quantum numbers were inferred
most probably $J^{PC}=1^{++}$, while the masses of the corresponding
charmonium states in the quark model $2^3P_1$ or $3^3P_1$ are hundreds
MeV above $M_X$. Recently, a study of the $\pi^+\pi^-\pi^0$ mass
distribution from the $X(3872)$ decay by the BABAR Collaboration
favors a negative parity assignment $2^{-+}$
\cite{delAmoSanchez:2010jr}.  However, still the mass of the
charmonium candidate in the quark model $1^1D_2$ disagrees with $M_X$.

Since it is difficult to assign $X(3872)$ to any conventional
$c\bar{c}$ charmonium state in the quark model, other alternative
explanations prevail. Because $M_X$ is close to the $D^*\bar{D}$
threshold, $X(3872)$ was interpreted as a loosely bound molecular
$D^*\bar{D}$ state \cite{Tornqvist:2004qy, Liu:2008tn, Thomas:2008ja,
  Ding:2009vj} or such a molecular state with some admixtures of $\rho
J/\psi$ and $\omega J/\psi$ \cite{Swanson:2003tb}.  It was also
suggested as a tetra quark state dominated by a diquark-antidiquark
structure \cite{Maiani:2004vq} or a $c\bar{c}g$ hybrid state
\cite{Li:2004sta}.  The proximity of the $X(3872)$ to $D^*\bar{D}$ threshold
may also imply that the cusp scenario should be treated seriously
\cite{Bugg:2004rk}.

The $D^{(*)}\bar{D}^{(*)}$ molecular states were proposed many years
ago\cite{DeRujula:1976qd, Tornqvist:1991ks}.  To bind the two mesons
together in a molecular state, additional interaction beyond the quark
potential model should be introduced between the two mesons. Naturally
the meson exchanges were first considered.  The inter-meson potentials
from light meson exchanges can be easily treated in the frame of the
heavy quark effecitive theory (HQET). The studies showed that the
cutoff parameters in the form factors are critical to the formation of
the molecular states \cite{Suzuki:2005ha, Meng:2005er, Zhu:2007wz,
  Liu:2008tn}.  Other interactions such as the gluon exchanges were
also considered \cite{Swanson:2003tb, Wong:2003xk, Wang:2009aw}.

Basically, the form factor is related to the inner quark wave
functions of the meson state.  To avoid the uncertainty from the form
factor, in this work we will investigate the possible molecular states
directly using the quark wave function within the quark model. Our
calculation is based on the multi-Gaussian function expansion of the
quark wave function of the molecular state, which is a rather simple
and efficient variational method to study many-body ground bound
states \cite{Kameyama:1989zz, Varga:1996zz, Brink:1998as,
  Wang:2009aw}.

We will assume that the light mesons ($\pi$, $\eta$, $\rho$, $\omega$
and $\sigma$) are exchanged between the light quarks from different
hadrons. The mechanism can be understood from the chiral quark model
\cite{Manohar:1983md} which was proposed by
Weinberg\cite{Weinberg:1978kz} and formulated by Manohar and
Georgi\cite{Manohar:1983md}.  The key feature of the chiral quark
model is that between the QCD confining scale and the chiral symmetry
breaking scale, QCD can be roughly described by an effective
Lagrangian of quarks and pseudo-scalar mesons of Numbu-Goldstone
bosons. The strong interaction between hadrons is thus described
mainly by the exchange of pseudo-scalar mesons at long-range. In
practice, the chiral quark model can be further extended with more
mesons to account for the intermidiate-range interaction. The light vector
resonances $\rho$, $K^*$, etc. can be realized as the dynamical gauge
bosons of a hidden local symmetry \cite{Georgi:1989xy}. The scalar
meson $\sigma$ can be explicitly introduced from the linear
realization, as in the linear sigma model.  Chiral quark model was
widely used in the study of nuclear forces \cite{Zhang:1997ny,
  Dai:2003dz} and recently used to study the molecular states
\cite{Liu:2009zzf}.

In section 2, we will introduce the quark model with the meson
exchange interactions in our work.  In section 3, the multi-Gaussian
function expansion method and the configuration space of the molecular
states are presented. We will study both the $D^{(*)}\bar{D}^{(*)}$ and
$B^{(*)}\bar{B}^{(*)}$ molecular states. Finally we will give a short
summary.

\section{The quark model with meson exchange}

The Hamiltonian in a quark model can be written as
\begin{equation}
  H = \sum_{i}(m_i+\frac{P_i^2}{2m_i})
  -\frac{3}{4}\sum_{i<j}\left(F_i\cdot{F_j}V_{ij}^c
    +F_i\cdot{F_j}S_i\cdot{S_j}V_{ij}^s\right).
\end{equation}
The first part is the non-relativistic kinetic energy, where $m_i$'s
are the constituent quark masses. The second part is the central
potential, where $F_i=\frac{1}{2}\lambda_i$ are the well known
$SU(3)_c$ Gell-Mann matrices. Apart from a constant, the central
potential usually is a combination of the one gluon exchange coulomb
potential and the linear confinment:
\begin{equation}
V_{ij}^c = -\frac{\kappa}{r_{ij}}+\frac{r_{ij}}{a_0^2}-M_0,
\end{equation}
where $r_{ij} = |r_i-r_j|$ is the distance between quark $i$ and quark
$j$.  The last part is the color-magnetic interaction, where
$S_i$=$\frac{1}{2}\sigma_i$ are the Pauli matrices.

In our work we will use the Bhaduri quark model which is a rather
simple non-relativistic quark potential model \cite{Bhaduri:1981pn}.
The potential function $V_{ij}^s$ of color-magnetic interaction reads
\begin{equation}
  V_{ij}^s = \frac{4\kappa}{m_im_j}\frac{1}{r_0^2r_{ij}}e^{-r_{ij}/r_0}.
\end{equation}
The model parameters are
\begin{equation}
  \begin{aligned}
    &\kappa =102.67\text{MeV}, &&a_0 = 0.0326(\text{fm}/\text{MeV})^{\frac12},\\
    &M_0 = 913.5\text{MeV}, &&r_0 = 0.4545\text{fm},\\
    &m_u = m_d = 337\text{MeV}, &&m_s = 600\text{MeV},\\
    &m_c = 1870\text{MeV}, &&m_b = 5290\text{Mev}.
  \end{aligned}
\end{equation}

The quark model is very successful in describing the hadron properties
at low energy under the QCD confinemnet scale.  As is well known, due to
the color symmetry, the above quark potential does not provide direct
interaction between hadrons.  Here we will consider the exchange of
the light mesons between the two light quarks from different hadrons
based on the chiral quark model. At present, we will consider the
exchanges of $\pi$, $\eta$, $\omega$, $\rho$, and $\sigma$ mesons
only. The interaction Lagrangian densities are
\begin{enumerate}
\item Pseudoscalar
  \begin{equation}
    \mathcal{L}_{p}=i{g_{p}}\bar{\psi}(x){\gamma_{5}}\psi(x)\varphi(x),
  \end{equation}
\item Scalar
  \begin{equation}
    \mathcal{L}_{s}=g_s\bar{\psi}(x)\psi(x)\phi(x),
  \end{equation}
\item Vector
  \begin{eqnarray}
    \mathcal{L}_{v}&=&g_v\bar{\psi}(x)\gamma_\mu\psi(x)v^\mu(x)\nonumber\\
    &&+\frac{f_v}{2m_q}\bar{\psi}(x)\sigma_{\mu\nu}\psi(x)\partial^{\mu}v^{\nu}.
  \end{eqnarray}
\end{enumerate}
Here $m_q$ is the constituent quark mass, $\psi(x)$ is the constituent
quark Dirac spinor field. $\varphi(x)$, $\phi(x)$, $v^\mu(x)$ are the
pseudoscalar, scalar and vector intermediate meson fields
respectively. In the case of isovector mesons, the above meson fields
should be replaced by $\bm\tau\cdot\bm\varphi$, $\bm\tau\cdot\bm\phi$
or $\bm\tau\cdot\bm{v}^\mu$ respectively, where $\bm{\tau}$ are the
iso-spin Pauli matrices. From the effective Lagrangian we can obtain
the effective potential between two quarks. In the coordinate space
\cite{Machleidt:1987hj},
\end{multicols}
\ruleup
\begin{enumerate}
\item Pseudoscalar
  \begin{equation}{\label{quarkpotentialps}}
    V_p(r)=\frac{g_p^2}{48\pi}m_p
    \left(\frac{m_p}{m_q}\right)^2Y_1(m_pr)
    \bm{\sigma}_1\cdot\bm{\sigma}_2 ,
  \end{equation}
\item Scalar
  \begin{equation}{\label{quarkpotentialscal}}
    V_s(r)=-\frac{g_s^2}{4\pi} m_s
    \left[ Y(m_sr)- \left(\frac{m_s}{m_q}\right)^2  \frac14 Y_1(m_sr)
      \right] ,
  \end{equation}  
\item Vector
  \begin{align}{\label{eq:quarkpotentialvec}}
    V_v(r)=&\frac{g_v^2}{4\pi}m_v\left\{Y(m_vr)
    +\left(\frac{m_v}{m_q}\right)^2 \frac12 Y_1(m_vr)
    \right\}
    +\frac{g_vf_v}{4\pi}m_v
    \left(\frac{m_v}{m_q}\right)^2 \left[ \frac12 Y_1(m_vr)
      +\frac{1}{3} Y_1(m_vr) \bm{\sigma}_1\cdot\bm{\sigma}_2
    \right]\nonumber\\
    &+\frac{f_v^2}{4\pi}m_v
    \left(\frac{m_v}{m_q}\right)^2 \frac16 Y_1(m_vr)
    \bm{\sigma}_1\cdot\bm{\sigma}_2 .
  \end{align}
\end{enumerate}
\ruledown \vspace{0.5cm}
\begin{multicols}{2}
We have neglected the momentum dependence of all potentials in the
present work. The spin-orbit interaction and the tensor-force are also
dropped out here since we only consider the molecular ground states.  The
functions $Y$, $Y_1$ are defined as follows:
\begin{align}
  Y(x)=&\frac{e^{-x}}{x}\\
  Y_1(x)=&\frac{e^{-x}}{x}- 4\pi\delta^3(\bm{x}) .
\end{align}
Note that on the quark level we need not introduce the form factors to
treat the $\delta$-interaction. The meson's quark wave functions will
naturally smear out the singularity.

The $q\bar{q}$ potentials are obtained from the above $qq$ potentials
by a $G$-parity transformation.  Also there should be an isospin factor
$\bm{\tau}_1\cdot\bm{\tau}_2$ in the above effective potential if an
isovector meson is exchanged.

From PDG\cite{Amsler:2008zzb}, the masses of exchanged mesons are
taken to be
\begin{align*}
  &m_\pi=134.98\MeV, &&m_\eta=547.85\MeV, \\
  &m_\rho=775.5\MeV, &&m_\omega=782.7\MeV, \\
  &m_\sigma=600.0\MeV.
\end{align*}
The masses of relevant heavy flavor mesons are:
\begin{align*}
  &m_{D^0}=1864.5\MeV, &&m_{D^\pm}=1869.3\MeV, \\
  &m_{D{*0}}=2006.7\MeV, &&m_{D^{*\pm}}=2010.0\MeV, \\
  &m_{B^0}=5279.5\MeV, &&m_{B^\pm}=5279.1\MeV, \\
  &m_{B*}=5325.1\MeV.
\end{align*}
The coupling constants are taken from Ref.~\cite{Ding:2009vj}:
\begin{align*}
  &g_\pi=2.74 &&g_{\eta}=2.05, &&g_{\sigma}=3.30, \\
  &g_{\rho}=3.46, &&f_{\rho}=1.45, \\
  &g_{\omega}=5.28, &&f_{\omega}=0,
\end{align*}
which are extracted from the meson-nucleon coupling constants in the
well-known Bonn model \cite{Machleidt:1987hj} using the single-quark
operator approximation \cite{Riska:2000gd}.

\section{The heavy flavor molecular states}

In this work, we consider the possible molecular states constructed
from the pseudo-scalar heavy mesons ($D$, $B$) and their vector
partners ($D^*$, $B^*$).  The states involving strange mesons are not
considered here. The corresponding charmed combinations are:
$D$-$\bar{D}$, $D^*$-$\bar{D^*}$, $D^*$-$\bar{D}$.  Since the charmed
mesons belong to the $I=\frac12$ representation of isospin $SU(2)$, 
the possible isospins of the $D^{(*)}$-$\bar{D}^{(*)}$ system are
$I=0$, $1$.  Following Ref.~\cite{Liu:2008fh}, we label the
$D$-$\bar{D}$, $D^*$-$\bar{D}$ and $D^*$-$\bar{D}^*$ systems as
$\Phi_{IJ}$, $\Phi^*_{IJ}$ and $\Phi^{**}_{IJ}$ respectively, while
the $D^*$-$\bar{D}$ system with negative charge conjugate ($C=-1$) is
labeled as $\hat\Phi^*_{IJ}$. Below we pick up the neutral
state to represent the isospin multiplet:
\begin{subequations}
\label{eq:hadronwf}
\begin{enumerate}
\item $I=0$
  \begin{align}
    \Phi^{0}_{00} &= \frac{1}{\sqrt{2}}\left( \bar{D}^0D^0-D^-D^+\right) \\
    \Phi^{**0}_{0J} &= \frac{1}{\sqrt{2}}
    \left( \bar{D}^{*0}D^{*0}-D^{*-}D^{*+}\right)_J \quad (J=0,1,2)\\
    \Phi^{*0}_{01} &= \frac12 [D^0\bar{D}^{*0}-D^{+}D^{*-}\nonumber\\
    &-C(\bar{D}^0D^{*0}-D^{-}D^{*+})],
  \end{align}
\item $I=1$
  \begin{align}
    \Phi^{0}_{10} &= \frac{1}{\sqrt{2}}\left( \bar{D}^0D^0+D^-D^+\right) \\
    \Phi^{**0}_{1J} &= \frac{1}{\sqrt{2}}
    \left( \bar{D}^{*0}D^{*0}+D^{*-}D^{*+}\right)_J \quad (J=0,1,2)\\
    \Phi^{*0}_{11} &= \frac12 [D^0\bar{D}^{*0}+D^{+}D^{*-}\nonumber\\
    &-C(\bar{D}^0D^{*0}+D^{-}D^{*+})].
  \end{align}
\end{enumerate}
\end{subequations}
The states of the $B^{(*)}-\bar{B}^{(*)}$ combinations are constructed
similarly but named after $\Omega$.

To calculate the molecular state, we use the Rayleigh-Ritz variation
principle. The test wave function is taken to be a series of Gaussian
functions with various widths \cite{Kameyama:1989zz, Varga:1996zz,
  Brink:1998as}.  In our case of the $Q\bar{q}q\bar{Q}$ molecular
state, the test wave function between the $Q\bar{q}$ and $q\bar{Q}$
mesons is expanded to \cite{Wang:2009aw}
\begin{eqnarray}
  \label{eq-var-4}
  \psi_{1234}(r_{12},r_{34},r_{1234}) &=& \sum_{i}
  \alpha_{1234}^i \psi_{12}(r_{12})\psi_{34}(r_{34})\nonumber\\
  &&\times\exp(-\beta_{1234}^i r_{1234}^2),
\end{eqnarray}
where $\bm{r}_1$, $\bm{r}_2$, $\bm{r}_3$ and $\bm{r}_4$ are the
coordinates of $Q$, $\bar{q}$, $q$ and $\bar{Q}$,
respectively. $\bm{r}_{ij} = \bm{r}_i -\bm{r}_j$.  $r_{1234}$ is the
distance between the two meson clusters
\begin{equation}
  \bm{r}_{1234} = \frac{m_Q\bm{r}_1+m_q\bm{r}_2}{m_Q+m_q}
  -\frac{m_q\bm{r}_3+m_Q\bm{r}_4}{m_q+m_Q}.
\end{equation}
Each of the meson wave functions $\psi_{ij}(r_{ij})$ is also taken to
be a Gaussian function series
\begin{equation}
  \label{eq-var-2}
  \psi_{ij}(r_{ij}) = \sum_k \alpha_{ij}^k
  \exp(-\beta_{ij}^k r_{ij}^2) .
\end{equation}

Numerically the above wave function is determined by the variational
method in two steps. First the meson wave function (\ref{eq-var-2}) is
determined from the potential quark model.  Then the wave function
(\ref{eq-var-4}) of the molecular state is obtained from the meson
exchange potentials with the meson wave functions $\psi_{ij}$ fixed.

To reduce the amount of computation, the parameters $\beta^i$ and
$\alpha^i$ in each Gaussian function series are determined also in two
steps by one-dimensional minimization.  We first search a central
$\beta$ value using a single Guassian function. Then a set
$\{\beta^i\}$ of $2N+1$ elements is generated by scaling the $\beta$
value up and down by a scale factor $s$ \cite{Brink:1998as}:
\begin{equation}
  \beta^i = \beta s^{i-N}
\end{equation}
where $i=0,1,...,2N$.

The bound energies and the mean squared radii (rms) $\langle
\bm{r}_{1234}^2 \rangle^{1/2}$ of the $D^{(*)}\bar{D}^{(*)}$ molecular states
are listed in Table~\ref{table1}.
\end{multicols}
\ruleup
  \begin{center}
    \tabcaption{\label{table1}%
    The bound energies of the $D^{(*)}\bar{D}^{(*)}$ molecular states.
    In calculation II, all the meson coupling constants  are reduced
    by a factor of $0.7$, except the $\pi$ meson.}
    \footnotesize
\begin{tabular*}{170mm}{@{\extracolsep{\fill}}c|c|d{3.2}d{3.2}d{3.2}d{3.2}d{3.2}}
      \hline\hline \multicolumn{2}{c|}{$T=0$}
      & \multicolumn{1}{c}{$\Phi_{00}$}
      & \multicolumn{1}{c}{$\Phi^{*}_{01}$}
      & \multicolumn{1}{c}{$\Phi^{**}_{00}$}
      & \multicolumn{1}{c}{$\Phi^{**}_{01}$}
      & \multicolumn{1}{c}{$\Phi^{**}_{02}$} \\\hline
      \multirow{2}{*}{rms(fm)} & I & 1.34 & 1.37 & 0.94 & 1.28 & 1.48 \\
      & II & 2.60 & 2.47 & - & 3.14 & 2.54 \\\hline
      \multirow{2}{*}{$E(\MeV)$}
      & I & -29.7 & -30.7 & -45.1 & -24.6 & -24.6 \\
      & II & -1.9 & -2.6 & - & -1.2 & -2.3 \\\hline
      \hline \multicolumn{2}{c|}{$T=1$}
      & \multicolumn{1}{c}{$\Phi_{10}$}
      & \multicolumn{1}{c}{$\hat\Phi^{*}_{11}$}
      & \multicolumn{1}{c}{$\Phi^{**}_{10}$}
      & \multicolumn{1}{c}{$\Phi^{**}_{11}$}
      & \multicolumn{1}{c}{$\Phi^{**}_{12}$} \\\hline
      \multirow{2}{*}{rms(fm)} & I & 1.24 & 1.12 & 1.04 & 1.19 & 1.71 \\
      & II & - & - & - & - & - \\\hline
      \multirow{2}{*}{$E(\MeV)$}
      & I & -13.7 & -18.3 & -22.6 & -15.6 & -5.7 \\
      & II & - & - & - & - & - \\\hline\hline
    \end{tabular*}
    \vspace{6mm}
  \end{center}
\ruledown
\begin{multicols}{2}
The calculation shows that the meson exchange interaction is strong
enough to bind the molecular states.  The typical $\Phi^*_{01}$ is the
candidate molecular state for the $X(3872)$ .  However the bound
energy of $\Phi^*_{01}$ is $30.7$ MeV, which is too larger than
what we expect, for $X(3872)$ should be a loose bound molecular state.

However, there is some uncertainty in the meson coupling constants on the
quark level.  In the Bonn model for neucleon interaction, the form
factors \cite{Machleidt:1987hj}
\begin{equation}
  F_\alpha(\bm{k}^2) = \left( \frac{\Lambda_\alpha^2-m_\alpha^2}
    {\Lambda_\alpha^2+\bm{k}^2} \right)^{n_\alpha}
\end{equation}
are also introduced in the description of the meson baryon
couplings. Clearly from the meson mass dependence in the form factor,
the effective meson coupling constants decrease as the mass of
intermediate mesons increase.

Next we try to decrease the coupling constants to the mesons $\sigma$,
$\rho$, $\omega$ and $\eta$ with heavier masses by a factor $\lambda$
following Ref~\cite{Ding:2009vj}. The numerical results for the case of
$\lambda = 0.7$ are shown in Table~\ref{table1} as calculation II. Now
the bound energy of $\Phi^*_{01}$ is only $2.6$ MeV and the rms is
$2.47$ fm, which meet the interpretation of $X(3872)$ as a loose bound
molecular state. Other possible molecular states left are iso-scalar
$\Phi_{00}(0^{++})$, $\Phi_{01}^{**}(1^{++})$ and
$\Phi_{02}^{**}(2^{++})$.

The similar calculation to $B^{(*)}\bar{B}^{(*)}$ molecular states are
shown in Table~\ref{table2}.
\end{multicols}
\ruleup
  \begin{center}
  \tabcaption{\label{table2}%
    The bound energies of $B^{(*)}\bar{B}^{(*)}$ molecular states.
    In calculation II, all the meson coupling constants except
    the $\pi$ meson are scaled by a factor of $0.7$.}
\footnotesize
\begin{tabular*}{170mm}{@{\extracolsep{\fill}}c|c|d{3.2}d{3.2}d{3.2}d{3.2}d{3.2}}
      \hline\hline \multicolumn{2}{c|}{$T=0$}
      & \multicolumn{1}{c}{$\Omega_{00}$}
      & \multicolumn{1}{c}{$\Omega^{*}_{01}$}
      & \multicolumn{1}{c}{$\Omega^{**}_{00}$}
      & \multicolumn{1}{c}{$\Omega^{**}_{01}$}
      & \multicolumn{1}{c}{$\Omega^{**}_{02}$} \\\hline
      \multirow{2}{*}{rms(fm)} & I & 1.13 & 1.16 & 0.65 & 1.00 & 1.21 \\
      & II & 1.39 & 1.32 & 1.11 & 1.38 & 1.42 \\\hline
      \multirow{2}{*}{$E(\MeV)$}
      & I &-46.0 & -49.2 & -81.0 & -44.6 & -44.0 \\
      & II & -13.1 & -17.3 & -11.7 & -9.5 & -15.3 \\\hline
      \hline \multicolumn{2}{c|}{$T=1$}
      & \multicolumn{1}{c}{$\Omega_{10}$}
      & \multicolumn{1}{c}{$\hat\Omega^{*}_{11}$}
      & \multicolumn{1}{c}{$\Omega^{**}_{10}$}
      & \multicolumn{1}{c}{$\Omega^{**}_{11}$}
      & \multicolumn{1}{c}{$\Omega^{**}_{12}$} \\\hline
      \multirow{2}{*}{rms(fm)} & I & 0.72 & 0.66 & 0.62 & 0.68 & 0.84 \\
      & II& 1.01 & 0.90 & 0.84 & 0.93 & 1.29 \\\hline
      \multirow{2}{*}{$E(\MeV)$}
      & I & -38.6 & -48.9 & -59.3 & -46.6 & -25.6 \\
      & II & -10.4 & -14.8 & -19.3 & -13.7 & -5.2 \\\hline\hline
    \end{tabular*}
     \vspace{6mm}
  \end{center}
\ruledown
\begin{multicols}{2}
As we expect, the bound energies of $B^{(*)}\bar{B}^{(*)}$ molecular states
become larger. Even when the coupling constants are weakened by the
scale factor $\lambda=0.7$ in calculation II, the meson exchange is
still strong enough to bind the molecular states for all
$B^{(*)}-\bar{B}^{(*)}$ combinations.

\section{Summary}

Based on the meson exchange interaction between light quarks, we have
investigated the heavy molecular states in quark model.  The molecular
states are described by the four quarks wave function expanded as a
series of Gaussian functions.  The numerical results show that the
light meson exchanges of $\pi$, $\eta$, $\rho$, $\omega$ and $\sigma$
between the light $u$, $d$ quarks are strong enough to bind the heavy
molecular states.

However, the bound energies of the molecular states are tens of MeV
(up to $80$MeV in $B^{(*)}\bar{B}^{(*)}$ cases) if we adopt the
meson-quark coupling constants from the meson-nucleon coupling
constants simply using the single-quark operator approximation. The
results are unreliable as the bound energies are somehow close to
$\Lambda_{\text{QCD}}$ while only the long-range meson exchanges are
considered in our calculation.

After we consider the uncertainty of the coupling
constants which are deduced from the Bonn potential of neucleon
interaction by decreasing the $\eta$, $\rho$, $\omega$ and $\sigma$
couplings by a factor of $0.7$, the $X(3872)$ is well interpreted as a
loose molecular $1^{++}$ state. The calculation also shows that other
possible molecular states such as $1^{++}$ and $2^{++}$ may exist on
the threshold of $D^*\bar{D}^*$ (The $0^{++}$ scalar state on the threshold
of $D\bar{D}$ is complicated from the scalar admixture). Since the bind
energies of $B^{(*)}\bar{B}^{(*)}$W molecular states increase with the the
mass increase of heavy favor, there are more such molecular states
near the $B^{(*)}\bar{B}^{(*)}$ thresholds as we expect.

The main uncertainty in the work is on the estimation of meson-quark
coupling constants. To obtain a set of reliable coupling constants, we
can use the model to study the nucleon-nucleon interaction in the
future.

\acknowledgments{
   We would like to thank professor Shi-Lin Zhu for useful discussions.
}

\end{multicols}
\vspace{-1mm}
\centerline{\rule{80mm}{0.1pt}}
\vspace{2mm}
\begin{multicols}{2}
\bibliography{refs}
\end{multicols}
\end{document}